\documentclass[aps,showpacs,twocolumn,amsmath]{revtex4}
\usepackage{graphicx}
\def\>{\rangle}
\def\<{\langle}

\begin{document}
\title{Coherence Control of Adiabatic Decoherence in a Three-level Atom with Lambda Configuration}
\author{Xiao-Shu Liu$^{1,2}$, Wu Re-bing $^{3}$, Yang Liu$^{1,2}$ , Jing Zhang$^{3}$ and Gui Lu Long$^{1,2,4}$\thanks{%
Corresponding author:gllong@tsinghua.edu.cn}}
\address{$^{1}$Department of Physics, Tsinghua University, Beijing 100084,
P. R. China\\
$^{2}$Key Laboratory For Quantum Information and Measurements,
Beijing 100084, P. R. China\\
 $^{3}$Department of Automation, Tsinghua University, Beijing
100084, P. R. China\\
$^4$ Center for Atomic and Molecular NanoSciences, Tsinghua
University, Beijing 100084, P. R. China\\}
\date{\today}
\begin{abstract}
In this paper, we study the suppression of adiabatic decoherence
  in a three-level atom with $\Lambda$ configuration using bang-bang control technique. We have given the
decoupling bang-bang operation group, and programmed a sequence of
periodic radio frequency twinborn pulses to realize the control
process. Moreover, we have studied the process with non-ideal
situation and established the condition for efficient suppression
of adiabatic decoherence.
\end{abstract}
 \maketitle

\section{ INTRODUCTION}
\label{s1}

In recent years,  there have been increased interests in the study
of three-level quantum systems, for example in quantum
cryptography \cite{ r1}, quantum communication \cite{r2}, logic
qubit encoding\cite{r3}, entanglement measures \cite{r4,r5,r6},
quantum control \cite{r7,r8,r9}, quantum computation \cite{r10}.
In these areas, people are always confronted with the obstacle of
decoherence, by which the superposition of the quantum states is
destructed during the system evolution.  Preserving coherence is
essential in quantum information processing, hence solutions must
be sought to dynamically suppress the decoherence effects.

Up to date, several  classes of schemes of decoherence control
have been proposed, for example, error-correcting,  error-avoiding
codes and dynamic decoupling technique. The error-correcting codes
\cite{r11,r12,r13,r14,r15,r16,r17}  use conditional feedback
control  to compensate the loss of information due to decoherence
or dissipation. Error-avoiding codes \cite{r18,r19,r20} decouples
the  interaction between the quantum system and the environment
exploiting the symmetry properties of the system and the
interaction.

Dynamical decoupling methods have been developed to control
decoherence\cite{r21,r22,r23,r24,r25,r26} for decades. Haeberlen
and Waugh pioneered the work of coherent averaging
effects\cite{r27} using tailored pulse sequence. It has been
developed into a solid decoupling and refocusing technique in
nuclear magnetic resonance (NMR) \cite{r23,r29}. Motivated by
these ideas, a "bang-bang" control theory \cite{r21,r22} has been
proposed to dynamically suppress the decoherence by repetitively
imposing a sequence of radio-frequency pulses on a single qubit.
This active dynamical control in the bang-bang limit proves a nice
tool for engineering the evolution of coupled quantum subsystems.
In this paper, we will apply the bang-bang control technique to
suppress the decoherence induced by pure dephasing in a
three-level atom with $\Lambda$ configuration.

\section{Problem formulation}
\label{s2}

\begin{figure}[!h]
\includegraphics[width=8cm,height=6cm,angle=0]{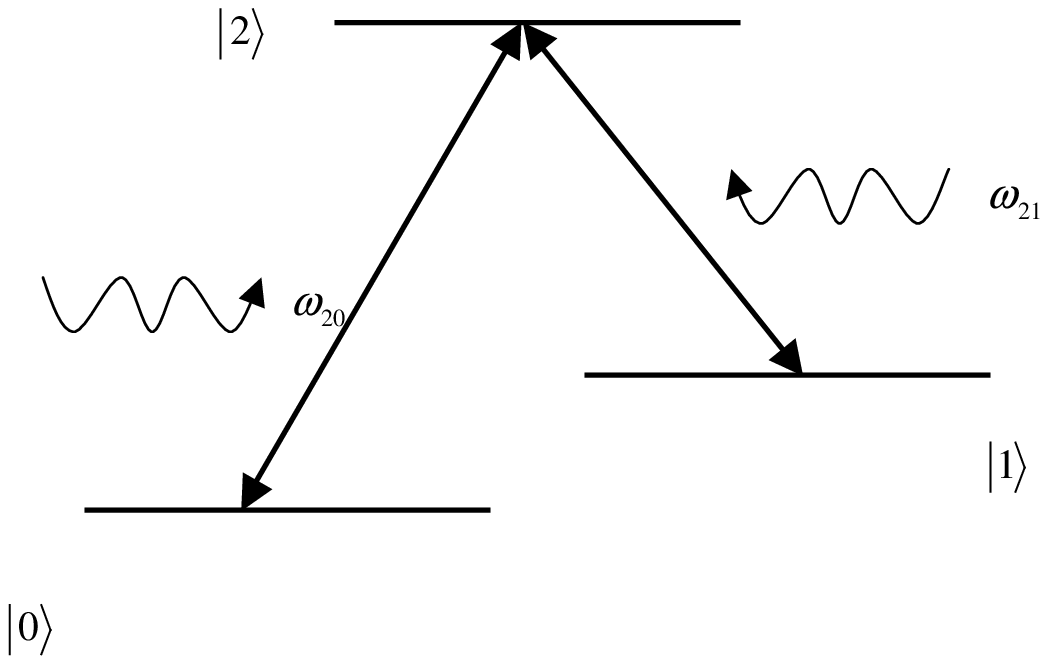}
\caption{Three-level atom in the $\Lambda$ configuration shooting
with two fields of frequencies $\omega_{20}$ and
$\omega_{21}$}\label{f1}
\end{figure}

The system we consider is a three-level atom with $\Lambda$
configuration under two resonant laser fields fields with frequencies
$$\omega_{20}=\frac{E_{2}-E_{0}}{\hbar},\,\,\omega_{21}=\frac{E_{2}-E_{1}}{\hbar},$$
respectively, as shown in Fig.\ref{f1}. Let $|0\>$, $|1\>$ and
$|2\>$ be the eigenstates of the unperturbed part of the
hamiltonian ${\cal H }_{0} $  of atom,  and the corresponding eigenvalues are $E_0$, $E_1$ and $E_2$ respectively
and  we assume $E_{0}<E_{1}<E_{2}$.
The two lower levels $|0\>$ and $|1\>$ are coupled
to a single upper level $|2\>$ in the $\Lambda$ type.

The total hamiltonian of the three-level atom can be expressed as
${\cal H}={\cal H }_{0}+{\cal H }_{R.F.}$, where
\begin{equation}
{\cal H}_{0}=E_{0}|0\>\<0|+E_{1}|1\>\<1|+E_{2}|2\>\<2| \label{e1}
\end{equation}
is the free hamiltonian and
\begin{widetext}
\begin{equation}
 {\cal H}_{R.F.}=-(g_{20}|2\>\<0|+g_{02}^{*}|0\>\<2|){\cal E}_{1}
 \cos \omega_{20} t-(g_{21}|2\>\<1|+g_{12}^{*}|1\>\<2|){\cal E}_{2}  \cos \omega_{21} t \label{e2}
\end{equation}
\end{widetext}
represents the interaction of the atom with the radiation fields.

Here we assume that the electric fields are linearly polarized
along the $x$-axis; $g_{ij}=g_{ji}^{*}=e\<i|x|j\>$ $(i,j=0,1,2)$
is the matrix element of the electric moment; ${\cal E}_{i}$
represents the amplitude of the electric field.

Before carrying out the calculation, we define some notation
 \begin{eqnarray}
&& \sigma_{z}^{(2,1)}\equiv|2\>\<2|-|1\>\<1|,\\
&&\sigma_{x}^{(2,1)}\equiv|2\>\<1|+|1\>\<2|,\\
&&\sigma_{y}^{(2,1)}\equiv i(|2\>\<1|-|1\>\<2|),\\
&&\sigma_{+}^{(2,1)}\equiv|2\>\<1|,\\
&&\sigma_{-}^{(2,1)}\equiv|1\>\<2|.
 \label{e3}
 \end{eqnarray}

The operators $\sigma_{x}^{(2,0)}$, $\sigma_{x}^{(1,0)}$,
$\sigma_{y}^{(1,0)}$, $\sigma_{y}^{(2,0)}$, $\sigma_{z}^{(2,0)}$
and $\sigma_{z}^{(1,0)}$ can be defined  similarly. With
these notations, the hamiltonian can be rewritten as
\begin{equation}
{\cal H}_{0}=\frac{\hbar \omega_{10}}{3}
\sigma_{z}^{(1,0)}+\frac{\hbar \omega_{20}}{3}
\sigma_{z}^{(2,0)}+\frac{\hbar \omega_{21}}{3}
\sigma_{z}^{(2,1)}+\frac{E_{0}+E_{1}+E_{2}}{3}. \label{e4}
\end{equation}
where
$$\omega_{10}=\frac{E_{1}-E_{0}}{\hbar},\omega_{20}=\frac{E_{2}-E_{0}}{\hbar},\omega_{21}=\frac{E_{2}-E_{1}}{\hbar}.$$
Usually, the constant energy $(E_{0}+E_{1}+E_{2})/3$ is
ignored.

For simplicity, we assume $g_{12}(=g_{21}^{*})$ and
$g_{10}(=g_{01}^{*})$ are real numbers. Then
\begin{eqnarray}
{\cal H}_{R.F.}&&=-g_{21}\cos \omega_{21}t
\sigma_{x}^{(2,1)}-g_{20}\cos \omega_{20}t\sigma_{x}^{(2,0)} \\
&&\equiv {\it u}_{1}(t)\sigma_{x}^{(2,0)}+  {\it
u}_{2}(t)\sigma_{x}^{(2,1)}.\label{e5}
\end{eqnarray}
Including the decoherence of the system due to the coupling to a
thermal reservoir, the total hamiltonian is
\begin{equation}
{\cal H}={\cal H}_{0}\otimes {\cal I}_{E}+{\cal I}_{S}\otimes
{\cal H}_{E}+{\cal H}_{\it SE}+{\cal H}_{R.F.}, \label{e7}
\end{equation}
where ${\cal H}_{0}$ and ${\cal H}_{R.F.}$ are the hamiltonians
described in (\ref{e4}) and (\ref{e5}). ${\cal H}_{E}$ and ${\cal
H}_{\it I}$ describe the internal hamiltonian of the environment
and its coupling hamiltonian to the three-level system.

The heat  bath is modelled as a large number of uncoupled bosonic modes,
namely a reservoir of simple harmonic oscillators with ground
state energy shifted to zero\cite{r23,r24,r25,r26,r27},
\begin{equation}
{\cal H}_{E}=\sum\limits_{k}\hbar\omega_{k}a_{k}^{\dagger}a_{k}.
 \label{e8}
\end{equation}
 The interaction hamiltonian is
expressed as
\begin{equation}
{\cal H}_{\it SE}=\hbar
\sum\limits_{k1}\sigma_{z}^{(2,0)}(g_{k1}a_{k1}^{\dagger}+g_{k1}^{*}a_{k1})+\hbar
\sum\limits_{k2}\sigma_{z}^{(2,1)}(g_{k2}a_{k2}^{\dagger}+g_{k2}^{*}a_{k2}).
\label{e9}
\end{equation}
where $g_{k1}$ and $g_{k2}$ are the  coupling constants
corresponding to the virtual exchanges of excitations with the
bath $|2\>\leftrightarrow |0\>$ and $|2\>\leftrightarrow|1\>$
transitions respectively.

Usually, we assume that the initial state of the total system is
disentangled, i.e.
$$\rho_{total}(0)=\rho_{S}(0)\otimes\rho_{E}(0),$$
and the thermal reservoir $\rho_{E}(0)$ is in thermal equilibrium
state that can be factorized into the tensor product of the
density operators of each mode
\begin{equation}
\rho_{E}=\prod\limits_{k}\theta_{k}\label{e10}
\end{equation}
where \begin{eqnarray*}
  Z_{k}&=&[1-\exp(-\frac{\hbar \omega_{k}}{k_{B}T})]\\
    \theta_{k}&=&Z_{k}^{-1}\exp(-\frac{\hbar\omega_{k}a_{k}^{\dagger}a_{k}}{k_{B}T}) \\
 \end{eqnarray*}
where $k_{B}$ is the Boltzmann constant and $T$ is the temperature
of the bath.

\section{Dynamical Suppression of Decoherence in a three-level atom in the ideal limits}
\label{s3}

Firstly, we sketch the main ideas of the dynamical decoupling
theory\cite{r29,r30,r31}. A bang-bang operation is  a
unitary operation that can be performed instantaneously, namely
 corresponding hamiltonian can be turned on for
negligible amounts of time $\tau$ with arbitrarily large strength.
Let $G_{B.B.}$ be the group consists of the implementable
bang-bang operations. The decoupling group ${\cal G}$ is defined
as a finite group of bang-bang decoupling operations, ${\cal
G}=\{g_{k}\}\subseteq G_{B.B.}$, where $k$ belongs to some finite
index set $K$. Then a decoupling-controller on ${\cal H}$ is
defined as the interactions of the system,  including a sequence of bang-bang operations
and free evolution.

Assume the cyclic time is $T_{c}$. Similar to the average
hamiltonian theory\cite{r26,r28}, we consider a given evolution
between the interval $(t-t_{0})=NT_{c}\equiv N|{\cal G}|\Delta t$
in the presence of decoupling-controller characterized by a
sequence of bang-bang operators $\{g_{k}\},(k=0,...,|{\cal
G}|-1)$. In a single cycle time $T_{c}$,
$U(T_{c})=\prod\limits_{k=0}^{|{\cal G}|-1}g_{k}^{+}U_{0}(\Delta
t)g_{k}\equiv e^{-i{\cal H}_{eff}T_{c}}$. In the ideal limit of
$T_{c}\rightarrow 0$ and $N\rightarrow \infty$, the effective
hamiltonian ${\cal H}$ approaches $${\cal H}_{eff}=\frac{1}{|{\cal
G}|}\sum\limits_{k=0}^{|{\cal G}|-1}g_{k}^{+}{\cal
H}g_{k}=\Pi_{{\cal G}}({\cal H}),$$
where
$\Pi_{{\cal G}}({\cal H})$ can be looked upon as a projector for operator
${\cal H}$ with the following properties\cite{r30}:
\begin{enumerate}
    \item projecting ${\cal H}$ into the centralizer $Z({\cal G})$: $Z({\cal
G})=\{X\in End({\cal H}_{S}^{space})|[X,g_{k}]=0,\forall
 k\}$;
    \item linearity:$\prod_{{\cal G}}({\cal H})=\Pi_{{\cal G}}
    ({\cal H}_{S}\otimes I_{E})+\Pi_{{\cal G}}(I_{S}\otimes {\cal H}_{E})+\Pi_{{\cal G}}({\cal H}_{SE})$.
\end{enumerate}
From the first property, the effective hamiltonian has a direct
symmetry characterization  $[{\cal H}_{eff},{\cal G}]=0$, and this
implies the composite system with
 decoherence is symmetrized by the group ${\cal G}$, i.e.
 all the components of the dynamics generated by ${\cal H}$, which are not invariant under the group
${\cal G}$, can be filtered out from the system's dynamics. When
$\Pi_{{\cal G}}({\cal H}_{SE})=0$, one can find from the second
property that the decoherence dynamics induced by the interaction
between the system and its environment has been averaged out.
Therefore, we can make use of this novel property to design
bang-bang control schemes with the dynamical decoupling group
${\cal G}$ to suppress the decoherence.

Applying the above ideas to the three-level atom with $\Lambda$
configuration with a dephasing interactions  characterized by Eq. (\ref{e9}), we
have found by trials and errors one decoupling group ${\cal
G}=\{I,h_{1},h_{2}\}$, where
\begin{equation}I=\left(
\begin{array}{ccc}1&0&0\\0&1&0\\0&0&1\end{array}\right), h_{1}=\left(
\begin{array}{ccc}0&0&i\\-1&0&0\\0&i&0\end{array}\right), h_{2}=\left(
\begin{array}{ccc}0&-1&0\\0&0&-i\\-i&0&0\end{array}\right),\label{e15}
\end{equation}
and they satisfy the following symmetrization equations
\begin{equation}\frac{1}{3}(\sigma_{z}^{(2,0)}+h_{1}^{\dagger}\sigma_{z}^{(2,0)}h_{1}+h_{2}^{\dagger}\sigma_{z}^{(2,0)}h_{2})=0,\label{e16}
\end{equation}
and
\begin{equation}\frac{1}{3}(\sigma_{z}^{(2,1)}+h_{1}^{\dagger}\sigma_{z}^{(2,1)}h_{1}+h_{2}^{\dagger}\sigma_{z}^{(2,1)}h_{2})=0.\label{e17}
\end{equation}

It is very interesting to note that these bang-bang operations can
be decomposed into
\begin{widetext}
\begin{equation}h_{1}=\left(
\begin{array}{ccc}1&0&0\\0&0&i\\0&i&0\end{array}\right)\left(
\begin{array}{ccc}0&0&i\\0&1&0\\i&0&0\end{array}\right)=\exp(i\frac{\pi}{2}\sigma_{x}^{(2,1)})\exp(i\frac{\pi}{2}\sigma_{x}^{(2,0)}),\label{e13}
\end{equation}
\end{widetext}
and
\begin{widetext}
\begin{equation}h_{2}=\left(
\begin{array}{ccc}0&0&-i\\0&1&0\\-i&0&0\end{array}\right)\left(
\begin{array}{ccc}1&0&0\\0&0&-i\\0&-i&0\end{array}\right)=\exp(-i\frac{\pi}{2}\sigma_{x}^{(2,0)})\exp(-i\frac{\pi}{2}\sigma_{x}^{(2,1)}).\label{e14}
\end{equation}
\end{widetext}
From these expressions, it is immediate to design a physical
realization for these operations using two R.F. pulses as given in
Eq. (\ref{e5}) with appropriate frequencies  that interact with
state transitions $|2\> \leftrightarrow |0\>$ and $|2\>
\leftrightarrow |1\>$ respectively. For example, $h_{1}$ can be
realized by two twinborn pulses, i.e. applying a $\pi$-pulse with
frequency $\omega _{20}$ at first and followed by another
$\pi$-pulse with frequency $\omega _{21}$. $h_{2}$ is realized
similarly.

\begin{figure}[!h]
\includegraphics[width=8cm,height=5cm,angle=0]{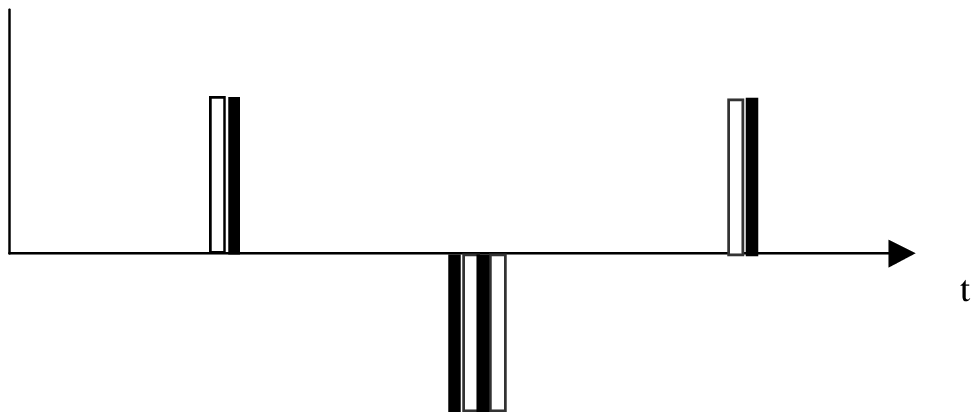}
\caption{A sequence of twinborn pulse operating in a cycle on the
three-level atom. The solid line represents the pulses with
frequency of $\omega_{21}$, and the hollow line represents the
pulses with frequency of $\omega_{20}$; $\pi$ pulses are on the
upside and $-\pi$ pulses are on the downside }\label{f2}
\end{figure}

With the above results, we can design a procedure to effectively
suppress the adiabatic decoherence with a sequence of periodic
twinborn pulses. In an elementary cycle, the twinborn-pulse
sequence is $\{h_{1},h_{1}^{\dagger},h_{2},h_{2}^{\dagger}\}$, as
shown in Fig. \ref{f2}.

In the first half of a cycle, system evolves under ${\cal H}={\cal
H}_{0}+{\cal H}_{E}+{\cal H}_{I} $ during $t_{0}\leq t \leq
t_{P}^{(1)}\equiv t_{0}+\Delta t $; at time $t_{P}^{(1)}$
twinborn-pulse $h_{1}$ is applied; after $2\tau_{P}$ units of
time, the pulse is switched off; then the system is governed by
${\cal H}$ during $t_{P}^{(1)}+2\tau_{P}\leq t\leq
t_{P}^{(2)}\equiv t_{0}+2\Delta t$, where $\tau_{P}$ is the pulse
width of each sub-pulse of the twinborn-pulse. In the second half
of the cycle, the twinborn-pulses $h_{1}^{\dagger}$ and $h_{2}$ is
applied and at time $t_{P}^{(2)}$ and $t_{P}^{(2)}+2\tau_{P}$
respectively; after another $2\tau_{P}$ units of time, the pulse
is switched off and the system evolves freely under ${\cal H}$
during $t_{P}^{(2)}+4\tau_{P}\leq t\leq t_{P}^{(3)}\equiv
t_{0}+3\Delta t$. At time $t_{P}^{(3)}$ the twinborn-pulses of
$h_{2}^{\dagger}$ begin. These complete a cycle. By repeating such
sequence of elementary cycles, one can suppress the adiabatic
decoherence completely in the ideal limits of $T_{c}\rightarrow 0$
and $N\rightarrow \infty $.

\section{Dynamical Suppression of Decoherence in a three-level atom with nonideal conditions}
\label{s4}

In last section, it is shown that the decoherence can be
completely removed in the ideal limits of $T_{c}\rightarrow 0$ and
$N\rightarrow \infty $. However, the ideal limits that
$T_{c}\rightarrow 0$ and $N\rightarrow \infty$ cannot be exactly
fulfilled in reality. In this section, we will give a quantitative
analysis of  effect of finite width  and finite-amplitude pulses
on the decoherence suppression.

The problem can be reformulated  in the interaction picture. Let
${\cal H}^{0}={\cal H}_{0}+{\cal H}_{E}$, then under the standard
state transformation $\exp(-i{\cal H}^{0}t)$, the interaction
${\cal {H}}_{SE}$ reads
\begin{eqnarray} \tilde{{\cal H}}_{\it SE}=\hbar
\sum\limits_{k1}\sigma_{z}^{(2,0)}(g_{k1}a_{k1}^{\dagger}e^{i\omega_{k1}t}+g_{k1}^{*}a_{k1}e^{-i\omega_{k1}t})\nonumber\\
 +\hbar
\sum\limits_{k2}\sigma_{z}^{(2,1)}(g_{k2}a_{k2}^{\dagger}e^{i\omega_{k2}t}+g_{k2}^{*}a_{k2}e^{-i\omega_{k2}t}),
\label{e24}
\end{eqnarray}
and the free unitary evolution of the composite system is
\begin{eqnarray}
\tilde{U}(t_{0},t)=\exp^{
\frac{\sigma_{z}^{(2,0)}}{2}\sum\limits_{k1}[a_{k1}^{\dagger}e^{i\omega_{k1}t_{0}}\xi_{k1}(t-t_{0})-h.c.]}\nonumber\\
\times \exp^{
\frac{\sigma_{z}^{(2,1)}}{2}\sum\limits_{k2}[a_{k2}^{\dagger}e^{i\omega_{k2}t_{0}}\xi_{k2}(t-t_{0})-h.c.]
 } \label{e24}
\end{eqnarray}
where
\begin{eqnarray}
\xi_{k1}(\Delta
t)&=&\frac{2g_{k1}}{\omega{k1}}(1-e^{i\omega_{k1}\Delta t})\nonumber\\
\xi_{k2}(\Delta
t)&=&\frac{2g_{k2}}{\omega{k2}}(1-e^{i\omega_{k2}\Delta t}).\nonumber
\end{eqnarray}

During an elementary cycle between time $t_{0}$ and
$t_{1}=t_{0}+3\Delta t+2\tau_{p}$, the state propagator can be
written as
\begin{widetext}
\begin{equation}
\tilde{U}_{P}(t_{0},t_{1})=\tilde{U}_{P4}\tilde{U}(t_{P}^{(2)}+4\tau
_{P},t_{P}^{(3)})\tilde{U}_{P3}\tilde{U}_{P2}\tilde{U}(t_{P}^{(1)}+2\tau
_{P},t_{P}^{(2)})\tilde{U}_{P1}\tilde{U}(t_{0}^{(1)},t_{P}^{(1)}),
 \label{e25}
\end{equation}
\end{widetext}
where \begin{eqnarray} \tilde{U}_{P1}&&= e^{i{\cal H}^{0}(t_{P}^{(1)}+\tau_{P})}
e^{i\frac{\pi}{2}\sigma_{x}^{(2,1)}}e^{-i{\cal H}^{0}(t_{P}^{(1)}+\tau_{P})}\nonumber\\
&&\times e^{i{\cal H}^{0}(t_{P}^{(1)}}
e^{i\frac{\pi}{2}\sigma_{x}^{(2,0)}}e^{-i{\cal H}^{0}(t_{P}^{(1)})}\nonumber\\
&&\approx e^{i{\cal H}^{0}(t_{P}^{(1)})}h_{1}e^{-i{\cal
H}^{0}(t_{P}^{(1)})}
 \label{e26}
\end{eqnarray}
and likewise,
\begin{eqnarray} \tilde{U}_{P2}\approx e^{i{\cal H}^{0}(t_{P}^{(2)})}h_{1}^{\dagger}
e^{-i{\cal H}^{0}(t_{P}^{(2)})}\nonumber\\
\tilde{U}_{P3}\approx e^{i{\cal H}^{0}(t_{P}^{(2)})}h_{2}e^{-i{\cal H}^{0}(t_{P}^{(2)})}\nonumber\\
\tilde{U}_{P4}\approx e^{i{\cal
H}^{0}(t_{P}^{(3)})}h_{2}^{\dagger}e^{-i{\cal
H}^{0}(t_{P}^{(3)})}.
 \label{e27}
\end{eqnarray}

Substituting Eqs.(\ref{e13},\ref{e14}) into Eq.(\ref{e27}), we
obtain
\begin{eqnarray}
\tilde{U}(t_{0},t)&=&e^{
\frac{\sigma_{z}^{(2,0)}}{2}\sum\limits_{k1}[a_{k1}^{\dagger}e^{i\omega_{k1}t_{0}}\xi_{k1}(\Delta t)-h.c.]}\nonumber\\
&&\times
e^{h_{1}^{\dagger}\frac{\sigma_{z}^{(2,0)}}{2}h_{1}\sum\limits_{k1}[a_{k1}^{\dagger}e^{i\omega_{k1}
t_{0}}e^{i\omega_{k1}\Delta t}\xi_{k1}(\Delta t)-h.c.]}\nonumber\\
&&\times
e^{h_{2}^{\dagger}\frac{\sigma_{z}^{(2,0)}}{2}h_{2}\sum\limits_{k1}[a_{k1}^{\dagger}e^{i\omega_{k1}t_{0}}
e^{i\omega_{k1}2\Delta
t}\xi_{k1}(\Delta t)-h.c.]}\nonumber\\
&& \times e^{
\frac{\sigma_{z}^{(2,1)}}{2}\sum\limits_{k2}[a_{k2}^{\dagger}e^{i\omega_{k2}t_{0}}\xi_{k2}(t-t_{0})-h.c.]}\nonumber\\
 &&\times e^{h_{1}^{\dagger}\frac{\sigma_{z}^{(2,1)}}{2}h_{1}\sum\limits_{k2}[a_{k2}^{\dagger}
 e^{i\omega_{k2}t_{0}}e^{i\omega_{k2}\Delta t}\xi_{k2}(\Delta
 t)-h.c.]}\nonumber\\
&&\times
e^{h_{2}^{\dagger}\frac{\sigma_{z}^{(2,1)}}{2}h_{2}\sum\limits_{k2}[a_{k2}^{\dagger}e^{i\omega_{k2}
t_{0}}e^{i\omega_{k2}2\Delta
t}\xi_{k1}(\Delta t)-h.c.]}\nonumber\\
&&\times e^{i{\cal H}_{0}(t_{1}-t_{0})}.
 \label{e28}
\end{eqnarray}

Imposing the above pulse sequences repeatedly, we then get the
general expression of the evolution under $N$ bang-bang control
cycles
$$\tilde{U}_{P}^{(N)}(t_{0},..,t_{N})=\tilde{U}_{P}(t_{N-1},t_{N})...\tilde{U}_{P}(t_{1},t_{2})
\tilde{U}_{P}(t_{0},t_{1}),
$$
where $t_n=t_0+3n\Delta t(n=1,..,N)$ is the ending time of the
$n$-th bang-bang control cycle.

With Eq.(\ref{e28}) and more careful calculations, we arrive at
\begin{eqnarray}
\tilde{U}_{P}^{(N)}&=&e^{
\frac{\sigma_{z}^{(2,0)}}{2}\sum\limits_{k1}[a_{k1}^{\dagger}e^{i\omega_{k1}t_{0}}f(n,N,\omega_{k1},\Delta
t)\xi_{k1}(\Delta t)-h.c.]}\nonumber\\
&&\times e^{-
\frac{\sigma_{z}^{(2,1)}}{2}\sum\limits_{k1}[a_{k1}^{\dagger}e^{i\omega_{k1}t_{0}}f(n,N,\omega_{k1},\Delta
t)\xi_{k1}(\Delta t)e^{i\omega_{k1}\Delta t}-h.c.]}\nonumber\\
&&\times e^{-
\frac{\sigma_{z}^{(1,0)}}{2}\sum\limits_{k1}[a_{k1}^{\dagger}e^{i\omega_{k1}t_{0}}f(n,N,\omega_{k1},\Delta
t)\xi_{k1}(\Delta t)e^{i\omega_{k1}2\Delta t}-h.c.]}\nonumber\\
&&\times e^{
\frac{\sigma_{z}^{(2,1)}}{2}\sum\limits_{k2}[a_{k2}^{\dagger}e^{i\omega_{k2}t_{0}}f(n,N,\omega_{k2},\Delta
t)\xi_{k2}(\Delta t)-h.c.]}\nonumber\\
&& \times e^{
\frac{\sigma_{z}^{(1,0)}}{2}\sum\limits_{k2}[a_{k2}^{\dagger}e^{i\omega_{k2}t_{0}}f(n,N,\omega_{k2},\Delta
t)\xi_{k2}(\Delta t)e^{i\omega_{k2}\Delta t}-h.c.]}\nonumber\\
&& \times e^{-
\frac{\sigma_{z}^{(2,0)}}{2}\sum\limits_{k2}[a_{k2}^{\dagger}e^{i\omega_{k2}t_{0}}f(n,N,\omega_{k2},\Delta
t)\xi_{k2}(\Delta t)e^{i\omega_{k2}2\Delta t}-h.c.]}\nonumber\\
&&\times e^{iN{\cal H}_{0}3\Delta t},
 \label{e29}
\end{eqnarray}
where $f(n,N,\omega,\Delta t)=\sum\limits_{n=1}^{N}
e^{3i(n-1)\omega \Delta t}.$

Now we can give a quantitative estimation of the decoherence rate
according to the time dependence of non-diagonal matrix elements
of the reduced density matrix of the three-level atom. For
example, the coherence between the level $|0\>$ and $|2\>$ is
represented by
\begin{widetext}
\begin{eqnarray*}\label{e30}
\tilde{\rho}_{02}^{S}(t)&=&{\rm
Tr}
_{E}\{\<0|\tilde{U}_{P}^{(N)}\tilde{\rho}^{S}(0)\otimes\tilde{\rho}_{E}(0)\tilde{U}_{P}^{(N)\dagger}|2\>\}
\\
&=&\tilde{\rho}_{02}^{S}(0)Tr_{E}\left[\tilde{\rho}_{E}(0)\exp\{\sum\limits_{k1}a_{k1}^{\dagger}e^{i\omega_{k1}t_{0}}\eta_{k1}(\Delta
t)+\sum\limits_{k2}a_{k2}^{\dagger}e^{i\omega_{k2}t_{0}}\eta_{k2}(\Delta
t)-h.c\}\right]\exp\{-i3N\omega_{20}\Delta t\}\\
&=&\tilde{\rho}_{02}^{S}(0)\exp\{-i3N\omega_{20}\Delta t-\Gamma(
k1,\Delta t)-\Gamma(k2,\Delta t)\},
\end{eqnarray*}
\end{widetext}
where $$\Gamma( k1,\Delta t)\equiv
\sum\limits_{k1}\frac{|\eta_{k1}(\Delta
t)|^{2}}{2}\coth(\frac{\omega_{k1}}{2T}),$$
$$\Gamma( k2,\Delta
t)\equiv \sum\limits_{k2}\frac{|\eta_{k2}(\Delta
t)|^{2}}{2}\coth(\frac{\omega_{k2}}{2T}),$$
and
$$\eta_{k1}(\Delta t)=-\frac{1}{2}f(n,N,\omega_{k1},\Delta
t)\xi_{k1}(\Delta t)(2-e^{i\omega_{k1}\Delta
t}-e^{i\omega_{k1}2\Delta t}),$$ and
$$\eta_{k2}(\Delta t)=-\frac{1}{2}f(n,N,\omega_{k2},\Delta
t)\xi_{k2}(\Delta t)(1+e^{i\omega_{k1}\Delta
t}-2e^{i\omega_{k1}2\Delta t}).$$

On the other hand, the density matrix without the bang-bang
control is
\begin{equation}
\tilde{\rho}
_{02}^{SW}(t)=\tilde{\rho}_{02}^{S}(0)e^{e^{-i3N\omega_{20}\Delta
t}-\Gamma^{'}( k1,\Delta t)-\Gamma^{'}(k2,\Delta t)},
 \label{e31}
\end{equation}
where
\begin{eqnarray*}
  \Gamma^{'}( k1,\Delta t)&\equiv&
\sum\limits_{k1}\frac{|\xi_{k1}(N,\Delta
t)|^{2}}{2}\coth(\frac{\omega_{k1}}{2T})\\
 \xi_{k1}(N,\Delta
t)&=&f(n,N,\omega_{k1},\Delta t)\xi_{k1}(\Delta t),
\end{eqnarray*}
and
\begin{eqnarray*}
  \Gamma^{'}( k2,\Delta t)&\equiv&
\sum\limits_{k2}\frac{|\xi_{k2}(N,\Delta
t)|^{2}}{2}\coth(\frac{\omega_{k2}}{2T})\\
 \xi_{k2}(N,\Delta
t)&=&f(n,N,\omega_{k2},\Delta t)\xi_{k2}(\Delta t).
\end{eqnarray*}

Physically, there exists a finite cutoff frequency of the
environment $\omega_{c}$\cite{r26,r31}. For a single mode of
frequency $\omega_{k}$, the time needed to produce appreciable
dephasing is $\tau_{k}=\omega_{k}^{-1}$, so $\tau_{c}\sim
\omega_{c}^{-1}$ sets the shortest time scale (or memory time) of
the environment. When $\omega_{k}\Delta
t\in[0,\arccos(\frac{3}{4})]$, we get that $\Gamma( k,\Delta
t)\leq \Gamma^{'}( k,\Delta t)$, which means in the quiet regime
$\Delta t\leq\tau_{c}\arccos(\frac{3}{4})$ the pulses will
effectively suppress the decoherence.

\begin{figure}
\begin{center}
\includegraphics[width=8cm,height=6cm,angle=0]{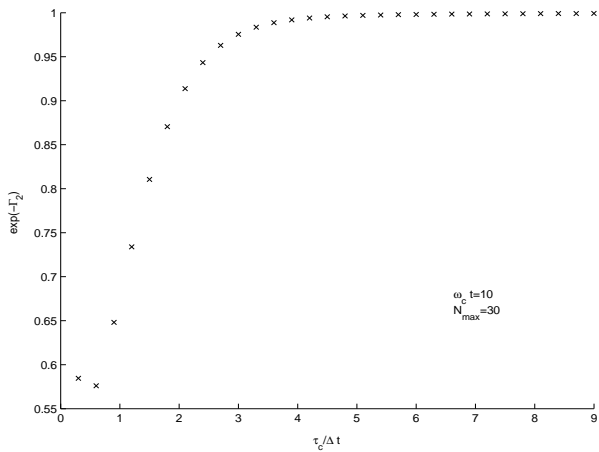}
\caption{Decoherence factor $\exp(-\Gamma_{2})$ in the presence of
periodical pulses at low-temperature, $\frac{\omega_{c}}{T}=100$.
Here $\alpha=0.25$. For a fixed time, each point corresponds to a
number of cycles, $N=1,..,N_{max}$. The maximum number of
bang-bang control cycles is $N_{max}=30$ for
$\omega_{c}t=10$.}\label{f3}
\end{center}
\end{figure}

In addition, $\Gamma( k,\Delta t)$ depends monotonously on the
cycle time $\Delta t$. In the ideal limit of $\Delta t\rightarrow
0(N\rightarrow \infty)$, the decoherence is completely suppressed
as a result of symmetrization.  To show this, we numerically
simulate one of the dephasing factors, $\Gamma( k2,\Delta t)$. In
the continuum limit of the bath mode, we can see
\begin{equation}
\Gamma_{2}\equiv\Gamma( k2,\Delta
t)=\int\limits_{0}^{+\infty}d\omega I(\omega)\times
\frac{|\eta_{k2}(\Delta t)|^{2}}{2}\coth \frac{\omega}{2T},
\end{equation}
where $I(\omega)=\frac{\alpha}{4}\omega
^{n}e^{-\omega/\omega_{c}}$, and $\alpha$ measures the strength of
the system-bath interaction and the index $n$ classifies different
environmental behaviors. For instance, the Ohmic environment
corresponds to $n=1$. From Fig. \ref{f3}, we can see that the
bigger the $N$(or smaller the $\Delta t$), the more effective the
bang-bang operation in suppressing the dephasing decoherence.

\section{Summary}
\label{s6}

In this paper, we have studied the suppression of adiabatic
decoherence of the three-level atom with $\Lambda$ configuration
using bang-bang control technique. The decoupling bang-bang
operation group is found, and the sequence of periodic R.F.
twinborn pulses is developed for the realization of the control
strategy. Moreover, we give a quantitative estimation of the
decoherence suppression in non-ideal limits. We also give the
condition of effectively suppressing this decoherence.

This work is supported by the National Fundamental Research
Program Grant No. 001CB309308, China National Natural Science
Foundation Grant No. 10325521, 60433050, 60074015,the Hang-Tian
Science Fund, and the SRFDP program of Education Ministry of
China.


\begin{thebibliography}{99}
\bibitem{r1} T. Durt, N. J. Cerf, N. Gisin and M. Zukowski,  {\it Phys. Rev. A. 67},
012311 (2003).

\bibitem{r2} C. Brukner, M. Zukowski, A. Zeilinger, {\it Phys. Rev. Lett. 89}, 197901
(2002).

\bibitem{r3} A. Grudka and A. Wojcik,  {\it Phy. Lett.} {\bf A 314}, 350 (2003).

\bibitem{r4} L-B. Fu, J-L. Chen, X-G. Zhao,
{\it Phys. Rev. A. 68}, 022323 (2003).

\bibitem{r5} J. L. Cereceda,
xxx.lanl.gov/quant-ph/0305043.

\bibitem{r6} H. Barnum, E. Knill, G. Ortiz, R. Somma and L. Viola, {\it Phys. Rev. Lett. 92}, 107902
(2004).

\bibitem{r7} U. Boscain, G. Charlot, J-P. Gauthier, {\it Optimal control of the Schr¡§odinger equation
with two or three levels}, in {\it  Nonlinear and adaptive control
(Sheffield, 2001)}, 33¨C43, Lecture Notes in Control and Inform.
Sci., 281, Springer, Berlin, 2003.

\bibitem{r8} D. D'Alessandro,
xxx.lanl.gov/quant-ph/0307129

\bibitem{r9}  Shlomo E. Sklarz and David J.Tannor,
arXiv:quant-ph/0402143 v1 19 Feb 2004.

\bibitem{r10} Kazuyuki Fujii , Kyoko Higashida , Ryosuke Kato , Yukako
Wada,  arXiv:quant-ph/0307066 v2 14 Jul 2003.

\bibitem{r11} A. R. Calderbank and P. W. Shor, {\it Phys. Rev. A.
54}, 1098 (1996); P. W. Shor,  {\it Phys. Rev. A. 52} R2493
(1995).

\bibitem{r12} R. Laflamme, C. Miquel, J. P. Paz and W. H. Zurek,
 {\it Phys. Rev. Lett. 77}, 198
(1996).

\bibitem{r13} W. H. Zurek and R. Laflamme, {\it Phys. Rev. Lett. 77}, 4683
(1996).

\bibitem{r14} D. Gottesman, {\it Phys. Rev. A. 54}, 1862
(1996).

\bibitem{r15} J. I. Cirac et al., {\it Science 273}, 1207
(1996).

\bibitem{r16} L. M. Duan and G. C. Guo, {\it Phys. Rev.
Lett. 79}, 1953(1997).

\bibitem{r17} S. Lloyd and  Slotine Jean-JacquesE, {\it Phys. Rev. Lett. 80},
4088 (1998).

\bibitem{r18} P. Zanardi and M. Rasetti, {\it Phy. Rev. Lett. 79},
3306 (1997).

\bibitem{r19} L. M. Duan and G. C. Guo {\it Phys. Rev. A. 57}, 2399
(1997).

\bibitem{r20} I. L. Chuang and Y. Yamamoto, {\it Phys. Rev. A. 52},
3489 (1995).

\bibitem{r21} L. Viola and S. Lloyd, {\it Phys. Rev. A. 58}, 2733
(1998).

\bibitem{r22} C. D' Helon, V. Protopopescu, and R. Perez,
 {\it J. Phys.} {\bf A 36}, 7129 (2003).


\bibitem{r23} D. Loss and D. P. DiVincenzo,  arXiv: cond-mat/0304118 .

\bibitem{r24} R. P. Feynman and A. R. Hibbs, Quantum Mechanics 8 and Path
Integrals (McGraw-Hill, NY, 1965).

\bibitem{r25} A. O. Caldeira and A. J. Leggett, {\it Phys. Rev. Lett.} {\bf 46},
211 (1981).

\bibitem{r26} A. J. Leggett, S. Chakravarty, A. T. Dorsey, M. P. A.
Fisher and W. Zwerger,{\it  Rev. Mod. Phys.} {\bf 59}, 1 (1987)

\bibitem{r27} S. Swain, {\it J. Phys.} {\bf A 5}, 1587 (1972.

\bibitem{r28} P. Zanardi, {\it Phy. Lett.} {\bf A 258}, 77 (1999).

\bibitem{r29} L. Viola, {\it Phy. Rev.} {\bf A 66}, 012307 (2002).

\bibitem{r30} L. Viola, S. Lloyd and E. Knill,{\it Phy. Rev. Lett.} {\bf 83},
4888 (1999).

\bibitem{r31} L. Viola,  E. Knill and S. Lloyd, {\it Phy. Rev. Lett.} {\bf 82},
2417 (1999).

\bibitem{r32}  U. Haeberlen and J.S. Waugh, {\it Phys. Rev.} {\bf 175}, 453, 1968.

\bibitem{r33}  U. Haeberlen, {\it High Resolution NMR in Solids: Selective
Averaging} (Academic Press, New York), 1976; R. R. Ernst, G.
Bodenhausen, and A. Wokaun, {\it Principles of Nuclear Magnetic
Resonance in One and Two Dimensions}, Oxford University Press,
Oxford, 1994.

\bibitem{r34}  D.G. Cory, M.D. Price, and T.F. Havel, {\it Physica} {\bf D 120}, 82
(1998); D.G. Cory et al., Fortschr. Phys. 48, 875 (2000).

\bibitem{r35} L. Viola and S. Lloyd, {\it Phys. Rev.} {\bf A 58}, 2733 (1998).

\end{thebibliography}
\end{document}